# Designing and using prior data in Ankylography: Recovering a 3D object from a single diffraction intensity pattern

E. Osherovich[1], O. Cohen[1], Y. C. Eldar[2], and M. Segev[1]

[1] Technion, Physics Department—Solid State Institute, Haifa, 32000, Israel
[2] Technion, Electrical Engineering Department, Haifa, 32000, Israel
**email**: oeli@tx.technion.ac.il

**Summary**

We present a novel method for Ankylography: three-dimensional structure reconstruction from a single shot diffraction intensity pattern. Our approach allows reconstruction of objects containing many more details than was ever demonstrated, in a faster and more accurate fashion

It was recently demonstrated [1] that a 3D object can be reconstructed from a single measurement of 2D diffraction intensity pattern by the technique called "Ankylography". However, "...rather than sparking a revoution in imaging, the idea has raised objections from researchers who say that it amounts to pulling a three-dimensional rabbit out of a two-dimensional hat" [2] (see also [3, 4]). To the best of our knowledge, the common consensus is that current reconstruction methods are not suitable for objects larger than about $15 \times 15 \times 15$ voxels (volume elements). This limitation stems from the fact that for an object of size $n \times n \times n$ voxels the number of measurements is proportional to $n^2$; hence, the ratio between the number of measurements and the number of unknowns is $1/n$, which makes the reconstruction problematic for large $n$. Furthermore, the current reconstruction methods do not easily allow introduction of *prior* information which can aid reconstruction. Here, we present a new approach enabling the reconstruction of objects with many more details, at very high accuracy.

Mathematically, Ankylography is closely related to the well-known problem of phase retrieval, where an object is reconstructed from the magnitude of its Fourier transform. In fact, the main difference between the standard phase retrieval and Ankylography is the reduced number of Ankylographic measurements that are only available on a spherical surface (Ewald's sphere) out of the full 3D cube, which is available in the standard phase retrieval. In this sense, Ankylography is close to our recent work on sub-wavelength coherent diffractive imaging [5]: both deals with object reconstruction from incomplete Fourier magnitude measurements. In [5] we demonstrated that the currently prevailing method for phase retrieval: the HIO algorithm [6], which was also used in [1] for Ankylography, is not suitable for reconstruction of objects from incomplete measurements. Instead, we developed a new reconstruction method [5] based on quasi-Newton optimization technique and apriori knowledge about the sought object (sparsity). Based on the similarity between these two problems, we embark upon a similar approach here: we use quasi-Newton optimization method and some *a priori* knowledge, as described below. Note that straightforward application of continuous optimization techniques, like quasi-Newton, has long been considered impractical for the phase retrieval problem. However, recent developments have paved the way for this approach when additional information is available [7]. One of such types of additional information is when a part of the sought object is known. This situation can be easily created by introducing some *known* object into the scene. This is what we do here: we assume prior knowledge of a small number of voxels that serve as sufficient constraints facilitating the full recovery of the rest of the sought 3D information. Physically, the known part of the object can be implemented by drilling tiny holes in the volume of the 3D medium we wish to image (see Figs. 1a and 1b). Modern fabrication techniques make it possible to realize such tiny holes at sufficiently high precision.

In the example below, the original object is comprised of the three most common atoms in biological molecules: carbon (C), hydrogen (H), and oxygen (O), which we model here as "billiard balls", each characterized by appropriate radius. That is, our *a priori* knowledge is that the object is comprised of three types of balls of known radii. We emphasize that we do not assume we know the number, the density, or the location of the balls. The original object is of size $128 \times 128 \times 128$ voxels constructed of 2C+3H+5O atoms (Fig.1a) placed randomly throughout the 3D volume. This structure (which simulates some unknown biological molecule) is "augmented" by four bars that constitute the known part of the combined object (Fig. 1b). Our reconstructed 3D objected is shown in Fig. 1d. Clearly, it is indistinguishable from the original object.

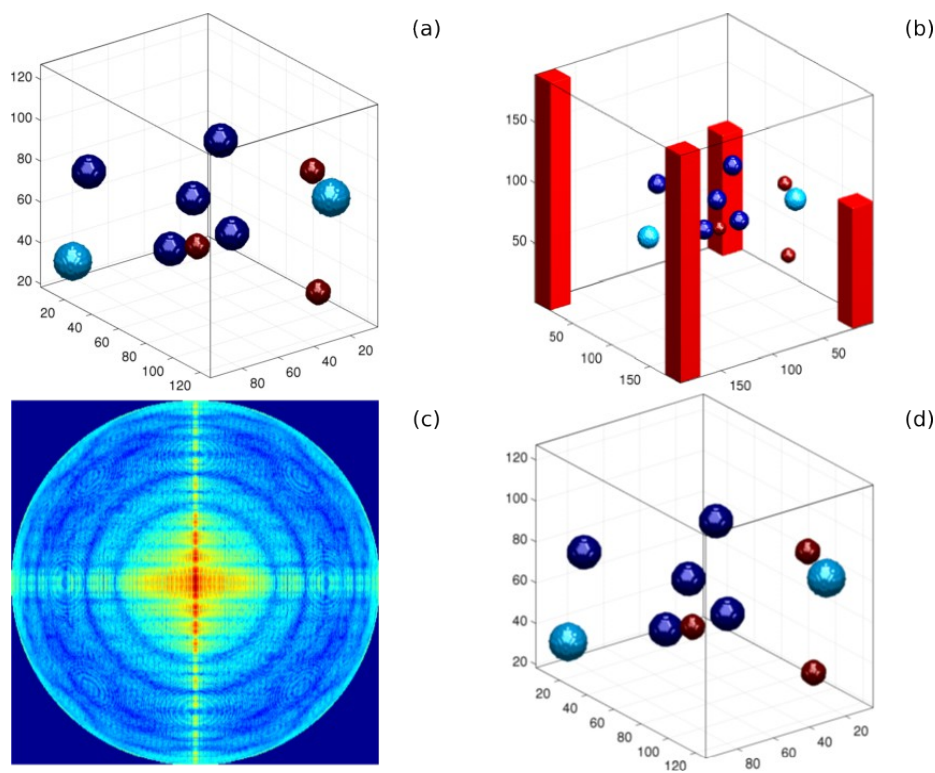

*Figure 1: Original object (a) and its "augmented" version with 4 holes drilled in the volume (b). Fourier domain measurements (flattened) (c). Our reconstruction (d).*

To conclude, we presented a new approach for Ankylography: the recovery of 3D objects from a single view. We believe that our approach will lead to much more powerful reconstruction techniques enabling the recovery of 3D objects containing much larger number of details than any existing algorithmic method, with the vision of visualizing the structure of 3D biological molecules from a single shot measurement.